\documentclass[12pt,a4paper,epsf]{article}
\usepackage{graphics}
\usepackage{amssymb,amsmath}
\usepackage[dvips]{lscape,graphicx}
\usepackage{cite}
\usepackage{longtable}
\usepackage{slashed}

\newcommand{\ct}{\cite}
\newcommand{\lb}{\label}

\newcommand{\bc}{\begin{center}}
\newcommand{\ec}{\end{center}}
\newcommand{\bd}{\begin{displaymath}}
\newcommand{\ed}{\end{displaymath}}
\newcommand{\be}{\begin{equation}}
\newcommand{\ee}{\end{equation}}
\newcommand{\ba}{\begin{array}}
\newcommand{\ea}{\end{array}}
\newcommand{\bea}{\begin{eqnarray}}
\newcommand{\eea}{\end{eqnarray}}
\newcommand{\bt}{\begin{tabular}}
\newcommand{\et}{\end{tabular}}

\newcommand{\bp}{\begin{picture}}
\newcommand{\ep}{\end{picture}}
\newcommand{\bfi}{\begin{figure}}
\newcommand{\efi}{\end{figure}}

\def\fun#1#2{\lower3.6pt\vbox{\baselineskip0pt\lineskip.9pt
\ialign{$\mathsurround=0pt#1\hfil##\hfil$\crcr#2\crcr\sim\crcr}}}

\begin{document}

\title{\LARGE \bf {New results at LHC confirming the vacuum stability and Multiple
Point Principle}}
\author{\large \bf L.V. Laperashvili ${}^{1}$\footnote
{laper@itep.ru},\; H.B. Nielsen
${}^{2}$\footnote{hbech@nbi.dk}\; and C.R. Das ${}^{3}$\footnote{das@theor.jinr.ru}\\\\
{\large \it ${}^{1}$ The Institute of Theoretical and
Experimental Physics,}\\
{\large\it National Research Center ``Kurchatov Institute'',}\\
{\large\it Bolshaya Cheremushkinskaya 25, 117218 Moscow, Russia}\\\\
{\large \it ${}^{2}$ Niels Bohr Institute,}\\
{\large \it Blegdamsvej, 17-21, DK 2100 Copenhagen, Denmark}\\\\
{\large \it ${}^{3}$ Bogoliubov Laboratory of Theoretical Physics}\\
{\large \it Joint Institute for Nuclear Research}\\
{\large \it International Intergovernmental Organization,}\\
{\large \it Joliot-Curie 6, 141980 Dubna, Moscow region, Russia}}

\date{}
\maketitle

\thispagestyle{empty}

\vspace{2cm}


%{\bf PACS:} 04.50.Kd, 98.80.Cq,
%12.10.-g, 95.35.+d, 95.36.+x

%\thispagestyle{empty}

%\clearpage \newpage

\begin{abstract}

In the present paper we argue that the correction to the Higgs
mass coming from the bound state of 6 top and 6 anti-top quarks,
predicted early by C.D.~Froggatt and ourselves, leads to the
Standard Model vacuum stability and confirms the accuracy of the
multiple point principle (principle of degenerate vacua) for all
experimentally valued parameters (Higgs mass, top-quark mass,
etc.). Fitting to get the vacuum degeneracy requires a mass of
the bound state, just in the region of the new two photon state in
LHC, 750-760 GeV.

\end{abstract}

\newpage

\section{Introduction}

In this paper we are concentrated on the vacuum stability problem
of the minimal Standard Model (SM) in the absence of new physics
at sub-Planckian energies. The value of the Higgs mass ($M_H =
125.66 \pm 0.34$ GeV) measured by ATLAS and CMS data \ct{1,2}, or
the more recent estimation from the combined ATLAS and CMS
analysis \ct{3,4}: $M_H = 125.09 \pm 0.24$ GeV, is intriguing: it
is quite close to the minimum $M_H$ value that ensures absolute
vacuum stability within the SM which, in turn, implies a vanishing
Higgs quartic coupling ($\lambda$) around the Planck scale.

Recently LHC data show hints of new resonances having invariant
masses of 300 and 750 GeV. Here, in this paper, we analyze the
problem of the vacuum stability/metastability in the SM in terms
of the new bosons, and try to show that these new resonances
discovered by LHC can explain the exact vacuum stability. This
problem is related with the concept of the Multiple Point 
Principle developed in 
Refs.~\ct{5mp,6mp,7mp,8mp,9mp,10mp,11mp,12mp,13mp,14mp,15mp,16mp,17mp}, 
presented in the book \ct{FN} and review \ct{DL}, and with the 
problem of tiny value of cosmological constant and dark energy 
(see for example Refs.~\ct{15mp,16mp,17mp,1S,2S,3S,4S,5S,6S}).

Our paper is organized as follows.

Section I is an Introduction. Here we have developed a concept of
the Multiple Point Principle (MPP) - theory of degenerate vacua
existing in Nature.

In general, a quantum field theory allows an existence of several
minima of the effective potential, which is a function of a scalar
field. If all vacua, corresponding to these minima, are
degenerate, having zero cosmological constants, then we can speak
about the existence of a multiple critical point (MCP) at the
phase diagram of theory considered for the investigation (see
Refs.~\ct{5mp,6mp,7mp,8mp,9mp}). In Ref.~\ct{5mp} Bennett and
Nielsen suggested the Multiple Point Model (MPM) of the Universe,
which contains simply the SM itself up to the scale $\sim 10^{18}$
GeV. They postulated a principle - Multiple Point Principle (MPP)
for many degenerate vacua - which should solve the finetuning
problem by actually making a rule for finetuning. If MPP is very
accurate we may have a new law of Nature, that can help us to
restrict coupling constants from a theoretical principle.

In Ref.~\ct{9mp} the MPP was applied (by the consideration of the
two degenerate vacua in the SM) for the prediction of the
top-quark and Higgs boson masses, which gave:
\be M_t = 173 \pm 5 \; {\rm GeV }, \qquad M_H = 135 \pm 9 \;
{\rm GeV }. \lb{1} \ee
In Ref.~\ct{16mp} it was argued that the exact degeneracy of vacua
in N = 1 supergravity can shed light on the smallness of the
cosmological constant. The presence of such vacua, which are
degenerate to very high accuracy, may also result in small values
of the quartic Higgs coupling and its beta function at the Planck
scale in the phase, in which we live.

Extending the Multiple Point Principal to yet another vacuum in
the pure SM, we have invented in Section 3 the idea that there
exists an exceptionally light scalar bound state $S$ constructed
by 6 top + 6 anti-top quarks, and that a boson condensate of these
New Bound States ($NBS$) should form a third phase of the SM
vacuum. According to the MPP, of course, this boson condensate
should get the coupling constants finetuned so as to have a vacuum
energy density (i.e. cosmological constant) very small, like the
other two vacua.

Section 2 considers a question: ``Could the Multiple Point Principle
be exact due to corrections from the new bound state 6 top + 6
anti-top?'' In previous Refs.~\ct{1nbs,2nbs,3nbs,4nbs} (see also
\ct{5nbs,6nbs,7nbs,8nbs,9nbs,10nbs,11nbs,12nbs,13nbs,14nbs}), we
and collaborators have speculated that 6 top + 6 anti-top quarks
should be so strongly bound that the bound states would
effectively function at low energies as elementary particles and
can be added into loop calculations as new elementary particles in
the theory, and seen, if produced, as resonances. But because of
their at the end composite nature they would deviate from a fully
fundamental particles by having formfactors cutting off their
interactions for high four momenta. The exceptional smallness of
the mass of the new bound state particle $S$ is in fact a
consequence of the degeneracy of the vacua, and thus of the
Multiple Point Principle.

In Section 3 we are concentrated on the vacuum stability problem
in the SM. In order to assess if the measured Higgs mass is
compatible with such a peculiar condition as an absolute vacuum
stability in the SM, a precise computation is needed. The study of
the stability of the SM vacuum has a long history
\ct{17vac,18vac,19vac,20vac,21vac,22vac,23vac,24vac,25vac,26vac,27vac,
28vac1,28vac2,29vac,30vac,31vac,32vac,33vac,34vac,35vac,36vac,37vac}. For
energies higher than electroweak scale the analysis of vacuum
stability is reduced to the study of the renormalization group
evolution of the Higgs quartic coupling $\lambda$. The prediction
(\ref{1}) for the mass of the Higgs boson was improved by the
calculation of the two-loop radiative corrections to the effective
Higgs potential \ct{Deg}. The prediction $129 \pm 2$ GeV in
Ref.~\ct{Deg} provided the possibility of the theoretical
explanation of the value $M_H \simeq 125.7$ GeV observed at the
LHC. The authors of Ref.~\ct{But} have shown that the most
interesting aspect of the measured value of $M_H$ is its
near-criticality. They extrapolated the SM parameters up to the
high (Planck) energies with full 3-loop NNLO RGE precision.

It was shown in Ref.~\ct{Deg} that the observed Higgs mass $M_H =
125.66 \pm 0.34$ leads to a negative value of the Higgs quartic
coupling $\lambda$ at some energy scale below the Planck scale,
making the Higgs potential unstable or metastable. Since the
measured value of $M_H$ is in a window of parameters where the SM
can be extrapolated till the Planck scale with no problem of
consistency but also that instability could arise, a highly
precise analysis for the vacuum stability becomes quite necessary.
With the inclusion of the three-loop RG equations (see \ct{But})
and two-loop matching conditions \ct{Deg}, the instability scale
occurs at $10^{11}$ GeV (well below the Planck scale) meaning that
at that scale the effective potential starts to be unbounded from
below or that a new minimum can appear, and there is a non-trivial
transition probability to that minimum. According to
Refs.~\ct{Deg,But}, the experimental value of the Higgs mass gives
scenarios which are at the border between the absolute stability
and metastability, the measured value of $M_H$ puts the SM in the
so-called near-critical position. Using the present experimental
uncertainties on the SM parameters (mostly the top quark mass)
Ref.~\ct{Deg} cannot conclusively establish the fate of the EW
vacuum, although metastability is preferred. The above statement
is the motivation for making a refined study of the vacuum
stability problem.

The careful evaluation of the Higgs effective potential by
Degrassi et al. \ct{Deg} combined with the experimentally measured
Higgs boson mass in the pure SM lead to the energy density getting
negative for high values of the Higgs field. E.g. the minimum in
this effective potential at the Higgs field being about $10^{18}$
GeV would have a negative energy density, or cosmological
constant, and formally the vacuum in which we live would be
unstable, although it is in reality just metastable with an
enormously long life time, if it is not deliberately made to
decay, what would be extremely difficult. It is however only
unstable vacuum with a very little margin in as far as the
experimental mass of 125.7 GeV is indeed very close to the mass
129.4 GeV, which according to the calculations of Degrassi et al.
\ct{Deg} would make the $10^{18}$ GeV Higgs field vacuum be
degenerate with the present one. In this sense Nature has in fact
chosen parameters very close to ones predicted by the ``Multiple
Point Principle'' developed in Refs.~\ct{5mp,6mp,7mp,8mp,9mp}.

Section 4 considers a model of $6t + 6\bar t$ bound states. We
considered the effect from the new bound states $6t+6\bar t$ on
the measured Higgs mass. We considered all Feynman diagrams which
give contributions of $S$-resonances to the renormalization group
evolution of the Higgs quartic coupling $\lambda$. Then we
calculated the main contribution of the $S$-resonance to
$\lambda$, and showed that the resonance with mass $m_S\approx
750$ GeV, having the radius $r_0 = b/m_t$ with $b\approx 2.35$,
gives such a positive contribution to $\lambda$ (equal to the
$\lambda_S\approx + 0.01$) which compensates the asymptotic value
of $\lambda\approx - 0.01$, earlier obtained by Ref.~\ct{Deg}. As
a result, this leads to the transformation of the metastability of
the EW vacuum to the stability.

Section 5 is devoted to the estimation of the radius of the new
bound state S of $6t + 6\bar t$. First we reviewed the results of
the mean field approximation obtained by Kuchiev, Flambaum,
Shuryak and Richard.

Finally, we have considered an alternative radius estimation for
the new bound state S, suggested by Froggatt and Nielsen, who have
used the ``eaten Higgs'' exchange corrections.

Section 7 contains Summary and Conclusions. Here we present an
explanation how the LHC new resonances which can be the earlier
predicted bound states of 6 top and 6 anti-top quarks can provide
the vacuum stability in the Standard Model confirming a high
accuracy of the Multiple Point Principle.

\section {Could the Multiple Point Principle be exact due to
corrections from the new bound state 6 top + 6 anti-top?}

The purpose of the present article is to estimate the corrections
from the NBS 6 top + 6 anti-top to the Higgs mass 129.4 GeV
predicted by Degrassi et al. \ct{Deg}, using the requirement of
the exact MPP. This actually can be done by identifying a barely
significant peak obtained at the LHC Run 2 with pp collisions at
energy $\sqrt s = 13 $ TeV \ct{ATLAS1,ATLAS2,CMS}, with our bound
state/resonance of 6 top and 6 anti-top. Run 2 LHC data show hints
of a new resonance in the diphoton distribution at an invariant
mass of 750 GeV. Since the peak, which we identify with our NBS,
corresponds to a mass of 750 GeV, it means that inserting into our
calculation of the correction to the predicted Higgs mass this
mass of 750 GeV, we can confirm the possible vacuum stability and
exact Multiple Point Principle.

\subsection{Search for either resonance in pp collision data at
$\sqrt s = 13$ TeV from the ATLAS detector}

The Higgs boson, H, offers a rich potential for new physics
searches. Recently in Refs.~\ct{ATLAS1,ATLAS2,CMS} the ATLAS and
CMS collaborations have presented the first data obtained at the
LHC Run 2 with pp collisions at energy $\sqrt s = 13 $ TeV. The
ATLAS collaboration has 3.2 $fb^{-1}$ of data and claims an
excess in the distribution of events containing two photons, at
the diphoton invariant mass $M \approx 750$ GeV with $3.9\sigma$
statistical significance. The ATLAS excess consists of about 14
events suggesting a best-fit width $\Gamma$ of about 45 GeV with
$\Gamma/M \approx 0.06$. The result is partially corroborated by
the CMS collaboration with integrated luminosity of 2.6
$fb^{-1}$, which has reported a mild excess of about
10 $\gamma\gamma$ events, peaked at 760 GeV. The best fit has a
narrow width and a local statistical significance of $2.6\sigma$.
Assuming a large width $\Gamma/M \approx 0.06$ the significance
decreases to $2.0\sigma$, corresponding to a cross section of
about 6 fb.

Fig.~1(a) presents searches for a new physics in high mass
diphoton events in proton-proton collisions obtained from
the combination of 8 TeV and 13 TeV results. ATLAS and
CMS Collaborations show a new resonance in the diphoton
distribution at an invariant mass of 750-760 GeV.

Ref.~\ct{ATLAS} (see Fig.~1(b)) presents searches for resonant and
non-resonant Higgs boson pair production using 20.3 ${\rm{
fb}^{-1}}$ of proton-proton collision data at $\sqrt s = 8$ TeV
generated by the Large Hadron Collider and recorded by the ATLAS
detector in 2012. A 95\% confidence level upper limit is placed on
the non-resonant production cross section at 2.2 pb, while the
expected limit is $1.0 \pm 0.5$ pb. The difference derives from a
small excess of events, corresponding to $2.4\sigma$. In the
search for a narrow resonance decaying to a pair of Higgs bosons,
the expected exclusion on the production cross section falls from
1.7 pb for a resonance at 260 GeV to 0.7 pb at 500 GeV. The
observed exclusion ranges from 0.7-3.5 pb. It is weaker than
expected for resonances below 350 GeV. It is not excluded that the
results show a resonance with mass $\approx 300$ GeV.

\section{Higgs mass and vacuum stability in the
Standard Model}

A theory of a single scalar field is given by the effective
potential $V_{eff}(\phi_c)$ which is a function of the classical
field $\phi_c$. In the loop expansion $V_{eff}$ is given by
\be V_{eff} = V^{(0)} + \sum_{n=1} V^{(n)}, \lb{1a} \ee
where $V^{(0)}$ is the tree-level potential of the SM.

\subsection{The tree-level Higgs potential}

The Higgs mechanism is the simplest mechanism leading to the
spontaneous symmetry breaking of a gauge theory. In the Standard
Model the breaking
\be
 SU(2)_L\times U(1)_Y \to U(1)_{em}, \lb{2a}
\ee
achieved by the Higgs mechanism, gives masses of the gauge bosons
$W^{\pm}$, $Z$, the Higgs boson and fermions $f$.

With one Higgs doublet of $SU(2)_L$, we have the following
tree--level Higgs potential:
\be V^{(0)} = - m^2 \Phi^{+}\Phi + \lambda(\Phi^{+}\Phi )^2.
 \lb{3a} \ee
The vacuum expectation value of $\Phi$ is:
\be
 \left\langle \Phi\right\rangle = \frac{1}{\sqrt 2}\left(
 \ba{c}
 0\\
 v
 \ea
 \right), \lb{4a} \ee
where
\be
 v = \sqrt{\frac{m^2}{\lambda}}\approx 246\; {\mbox{GeV}}.
 \lb{5a} \ee
Introducing a four-component real field $\phi$ by
\be
 \Phi^{+}\Phi = \frac{1}{2}\phi^2, \lb{6a} \ee
where
\be
 \phi^2 = \sum_{i=1}^4 \phi_i^2, \lb{7a} \ee
we have the following tree-level potential:
\be
 V^{(0)} = - \frac{1}{2} m^2 \phi^2 + \frac{1}{4} \lambda
 \phi^4. \lb{8a} \ee
As is well-known, this tree--level potential gives the masses of
the gauge bosons $W$ and $Z$, fermions with flavor $f$ and the
physical Higgs boson $H$, which are described by the VEV parameter
$v$:
\bea
M_W^2 &=& \frac{1}{4} g^2 v^2, \lb{9a}\\
M_Z^2 &=& \frac{1}{4} \left(g^2 + g'^2\right) v^2, \lb{10a}\\
M_f &=& \frac{1}{\sqrt 2} g_f v, \lb{11a}\\
M_H^2 &=& \lambda v^2, \lb{12a}
\eea
where $g_f$ is the Yukawa couplings of fermion with the flavor
$f$.

\subsection{Stability phase diagram}

The vast majority of the available experimental data is consistent
with the SM predictions. No sign of new physics has been detected.
Until now there is no evidence for the existence of any particles
other than those of the SM, or bound states composed of other
particles. All accelerator physics seems to fit well with the SM,
except for neutrino oscillations. These results caused a keen
interest in possibility of emergence of new physics only at very
high (Planck scale) energies, and generated a great attention to
the problem of the vacuum stability: whether the electroweak
vacuum is stable, unstable, or metastable
\ct{17vac,18vac,19vac,20vac,21vac,22vac,23vac,24vac,25vac,26vac,27vac,
28vac1,28vac2,29vac,30vac,31vac,32vac,33vac,34vac,35vac,36vac,37vac}. A
largely explored scenario assumes that new physics interactions
only appear at the Planck scale $M_{Pl}= 1.22\cdot 10^{19}$ GeV
\ct{1he,2he,3he,4he} and \ct{Deg,But}. According to this scenario,
we need the knowledge of the Higgs effective potential
$V_{eff}(\phi)$ up to very high values of $\phi$. The loop
corrections lead the $V_{eff}(\phi)$ to values of $\phi$ which are
much larger than v, the location of the electroweak (EW) minimum.
The effective Higgs potential develops a new minimum at $v_2 \gg
v$. The position of the second minimum depends on the SM
parameters, especially on the top and Higgs masses, $M_t$ and
$M_H$. It can be higher or lower than the EW one showing a stable
EW vacuum (in the first case), or metastable one (in the second
case). Then considering the lifetime $\tau$ of the false vacuum
(see Ref.~\ct{36vac}), and comparing it with the age of the
universe $T_U$, we see that, if $\tau$ is larger than $T_U$, then
our universe will be sitting on the metastable false vacuum, and
we deal with the scenario of metastability.

Usually the stability analysis is presented by stability diagram
in the plane ($M_H,\,M_t$). The standard results are given by
Refs.~\ct{1he,2he,3he,4he,Deg,But}, and the plot is shown in
Fig.~2. For our purposes we were guided by the ``phase diagram''
presented in Ref.~\ct{37vac}. As it was noted by authors of
Ref.~\ct{37vac}, strictly speaking, this is not a phase diagram,
but this expression is still used due to the common usage in
literature.

We see that the plane ($M_H,\,M_t$) is divided in Fig.~2 into
three different sectors: 1) An absolute stability region (cyan),
where $V_{eff}(v) < V_{eff} (v_2)$, 2) a metastability region
(yellow), where $V_{eff} (v_2) < V_{eff} (v)$, but $\tau > T_U$, and
3) an instability (green) region, where $V_{eff} (v_2) < V_{eff}
(v)$ and $\tau < T_U$. The stability line separates the stability
and the metastability regions, and corresponds to $M_t$ and $M_H$
obeying the condition $V_{eff} (v) = V_{eff} (v_2)$. The
instability line separates the metastability and the instability
regions. It corresponds to $M_t$ and $M_H$ for $\tau = T_U$.

In Fig.~2 the black dot indicates current experimental values
$M_H\simeq 125.7$ GeV \ct{3,4} and $M_t\simeq 173.34$ GeV
\ct{PDG,Mtexp}. It lies inside the metastability region, and could
reach and even cross the stability line within $3\sigma$. The
ellipses take into account $1\sigma,\, 2\sigma$ and $3\sigma$,
according to the current experimental errors. When the black dot
sits on the stability line, then this case is named ``critical'',
according to the MPP concept \ct{5mp,6mp,7mp,8mp,9mp}: the running
quartic coupling $\lambda$ and the corresponding beta-function
vanish at the Planck scale:
\be \lambda\left(M_{Pl}\right) \sim 0 \quad {\rm{and}}\quad
\beta\left(\lambda\left(M_{Pl}\right)\right)\sim 0. \lb{13a} \ee
Fig.~2 shows that the black dot, existing in the metastability
region, is close to the stability line, and the ``near-criticality''
\ct{But} can be considered as the most important information
obtained for the Higgs boson.

The Higgs inflation scenario developed in
Refs.~\ct{1Bez,2Bez,3Bez,4Bez,5Bez} confirms the realization of
the conditions (\ref{13a}).

\subsection{Two-loop corrections to the Higgs mass from the effective
potential}

Still neglecting new physics interactions at the Planck scale, we
can consider the Higgs effective potential $V_{eff}(\phi)$ for
large values of $\phi$ \ct{22vac} (see also \ct{9mp,12mp}):
\be V_{eff}(\phi)\simeq \frac 14 \lambda_{eff}(\phi) \phi^4.
\lb{14a} \ee
Here $V_{eff}(\phi)$ is the renormalization group improved (RGE)
Higgs potential \ct{22vac}, and $\lambda_{eff}(\phi)$ depends on
$\phi$ as the running quartic coupling $\lambda(\mu)$ depends on
the running scale $\mu$. Then we have the one-loop, two-loops or
three-loops expressions for $V_{eff}$. The corresponding up to
date Next-to-Next-to-Leading-Order (NNLO) results were published
in Refs.~\ct{Deg,But,1NLO,2NLO,6Bez}. For a large range of values
of $M_H$ and $M_t$, the Higgs effective potential has a minimum
(see also \ct{12mp}). If the point $\phi=\phi_0=v$ is a minimum of
the $V_{eff}(\phi)$ for a given couple ($M_H,\,M_t$), then from
Eq.~(\ref{14a}) the stability line corresponds to the conditions:
\be V_{eff} \left(\phi_0\right) = 0 \quad {\rm {and}}\quad
V'_{eff} \left(\phi_0\right) = 0. \lb{15a} \ee
In general, MPP predicts that:
\be V_{eff} \left(\phi_{min 1}\right) = V_{eff} \left(\phi_{min
2}\right) = 0, \lb{16a} \ee
\be V'_{eff} \left(\phi_{min 1}\right) = V'_{eff} \left(\phi_{min
2}\right) = 0, \lb{16b} \ee
where $\phi_{min 1}=v$ is the first EW vacuum, and $\phi_{min
2}=v_2$ is the second Planck scale vacuum.

The red solid line of Fig.~3 shows the running of the
$\lambda_{eff}(\phi)$ for $M_H\simeq 125.7$ and $M_t\simeq
171.43$, which just corresponds to the stability line, that is, to
the stable vacuum. In this case the minimum of the $V_{eff}(\phi)$
exists at the $\phi=\phi_0=v\simeq 2.22\cdot 10^{18}$ GeV, where
according to (\ref{15a}):
\be \lambda_{eff}\left(\phi_0\right)=0 \quad {\rm {and}}\quad
\beta\left(\lambda_{eff}\left(\phi_0\right)\right)=0. \lb{17a} \ee
But as it was shown in Refs.~\ct{Deg,But}, this case does not
correspond to current experimental values.

The relation between $\lambda$ and the Higgs mass is:
\be \lambda(\mu) = \frac{G_F}{\sqrt 2} M_H^2 + \Delta
\lambda(\mu), \lb{18a} \ee
where $G_F$ is the Fermi coupling. In Eq.~(\ref{18a}) $\Delta
\lambda(\mu)$ denotes corrections arising beyond the tree level
case.

Computing $\Delta \lambda(\mu)$ at the one loop level, using
two-loop beta functions for all the SM couplings, Ref.~\ct{Deg}
obtained the first complete NNLO evaluation of $\Delta
\lambda(\mu)$ for the two-loop QCD and Yukawa contribution to
$\Delta \lambda(\mu)$ in the SM with the electroweak gauge
couplings switched off.

In Fig.~3 blue lines (thick and dashed) present the RG evolution
of $\lambda(\mu)$ for current experimental values $M_H\simeq
125.7$ GeV \ct{3,4} and $M_t\simeq 173.34$ GeV \ct{Mtexp}, and for
$\alpha_s$ given by $\pm 3\sigma$. The thick blue line corresponds
to the central value of $\alpha_s = 0.1184$ and dashed blue lines
correspond to errors of $\alpha_s$ equal to $\pm 0.0007$. After
the rapid variation of $\lambda(\mu)$ around the weak scale shown
in Fig.~3, these corrections play a significant role in
determining the evolution of $\lambda$ at high energies. Absolute
stability of the Higgs potential is excluded by Ref.~\ct{Deg} at
98\% C.L. for $M_H < 126$ GeV. In Fig.~3 we see that
asymptotically $\lambda(\mu)$ does not reach zero, but approaches
to the negative value:
\be \lambda \to - (0.01 \pm 0.002), \lb{19a} \ee
indicating the metastability of the EW vacuum. According to
Ref.~\ct{Deg}, the stability line shown in Fig.~3 by the red thick
line corresponds to
\be M_H = 129.4 \pm 1.8\;  {\rm{GeV}}. \lb{20a} \ee
The aim of the present paper is to show that the stability line
{\em could} correspond to the current experimental values of the
SM parameters for $M_H\simeq 125.7$ given by LHC.

\section{The effect from the new bound states $6t+6\bar t$ on the measured
Higgs mass}

In Refs.~\ct{1nbs,2nbs,3nbs,4nbs} there was suggested the
existence of new bound states (NBS) of 6 top + 6 anti-top quarks
as so strongly bound systems that they effectively function as
elementary particles. Later this theory was developed in
Refs.~\ct{5nbs,6nbs,7nbs,8nbs,9nbs,10nbs,11nbs,12nbs,13nbs,14nbs}.

\subsection{New bound states}

In Ref.~\ct{1nbs} there was first assumed that\\\\$\bullet$ there
exists $1S$-bound state $6t+6\bar t$ - scalar particle and color singlet;\\\\
$\bullet$ that the forces responsible for the formation of these
bound states originate from the virtual exchanges of the Higgs
bosons between top(anti-top)-quarks;\\\\
$\bullet$ that these forces are so strong that they almost
compensate the mass of 12 top-anti-top quarks contained in these
bound states.

The explanation of the stability of the bound state $6t+6\bar t$
is given by the Pauli principle: top-quark has two spin and three
color degrees of freedom (total 6). By this reason, 6 quarks have
the maximal binding energy, and 6 pairs of $t\bar t$ in $1S$-wave
state create a long lived (almost stable) colorless scalar bound
state $S$. One could even suspect that not only this most
strongly bound state $S$ of $6t+6\bar t$, but also some excited
states exist. It is obvious that excited to a 2s or 2p state (in
the atomic physics terminology), scalar and vector particles
correspond to the more heavy bound states of the $6t + 6\bar t$,
etc. Also there exists a new bound state $ 6t + 5\bar t$, which is
a fermion similar to the quark of the 4th generation having
quantum numbers of top-quark.

These bound states are held together by exchange of the Higgs and
gluons between the top-quarks and anti-top-quarks as well as
between top and top and between anti-top and anti-top. The Higgs
field causes attraction between quark and quark as well as between
quark and anti-quark and between anti-quark and anti-quark, so the
more particles and/or anti-particles are being put together the
stronger they are bound. But now for fermions as top-quarks, the
Pauli principle prevents too many constituents being possible in
the lowest state of a Bohr atom constructed from different
top-quarks or anti-top-quarks surrounding (analogous to the
electrons in an atom) the ``whole system'', analogous to the
nucleus in the Bohr atom. Because the quark has three color states
and two spin states meaning six internal states there is in fact a
shell (as in the nuclear physics) with six top quarks and
similarly one for six anti-top quarks. Then we imagine that in the
most strongly bound state just this shell is filled and closed for
both top and anti-top. Like in nuclear physics where the closed
shell nuclei are the strongest bound, we consider this NBS 6 top +
6 anti-top as our favorite candidate for the most strongly bound
and thus the lightest bound state $S$. Then we expect that our
bound state $S$ is appreciably lighter than its natural scale of
12 times the top mass, which is about 2 TeV. So the mass of our
NBS $S$ should be small compared to 2 TeV. In recent papers
\ct{4nbs,11nbs} C.D.~Froggatt and H.B.~Nielsen estimated masses
smaller than 2 TeV, but the calculation were too detailed to be
trusted. It should be had in mind that such a calculation could
strongly enhance the reliability of our finetuning principle -
multiple point principle. From MPP it is possible to calculate the
top-Yukawa coupling. Ref.~\ct{11nbs} gave a prediction of the top
Yukawa coupling: $g_t= 1.02 \pm 14 \%$. Since the experimental top
Yukawa coupling is $g_t=0.935$, there is a hope that our model
could be right.

In this paper, taking into account that LHC recently discovered
new resonances \ct{ATLAS1,ATLAS2,CMS,ATLAS}, we analyze the
problem of the vacuum stability/metastability in the SM in terms
of these new bosons, and try to show that these new resonances can
explain the exact vacuum stability and the exact Multiple Point
Principle developed in
Refs.~\ct{5mp,6mp,7mp,8mp,9mp,10mp,11mp,12mp,13mp,14mp,15mp,16mp,17mp}.

\subsection{The main diagram correcting the effective Higgs
self-interaction coupling constant $\lambda$}

Estimating different contributions of the bound state $S$ (see
Figs.~4(a,b)) we found that the main Feynman diagram correcting
the effective Higgs self-interaction coupling constant
$\lambda(\mu)$ is the diagram of Fig.~4(a) containing the bound
state $S$ in the loop.

Now instead of Eq.~(\ref{18a}) we have:
\be \lambda(\mu) = \frac{G_F}{\sqrt 2} M_H^2 + \delta
\lambda(\mu) + \Delta \lambda(\mu), \lb{19a} \ee
where the term $\delta \lambda(\mu)$ denotes the loop corrections
to the Higgs mass arising from all NBS, and the main contribution
to $\delta \lambda(\mu)$ is the term $\lambda_S$, which
corresponds to the contribution of the diagram of Fig.~4(a):
\be \delta \lambda(\mu) = \lambda_S + ... \lb{20a} \ee
The rest (see Fig.~4(b)) can be at most of a similar order of
magnitude as the ``dominant'' one. But even if we should get a
factor 2 or so it would for the mass $m_S$ inside the quantity
going into the 4th power only mean a factor the fourth root of 2,
which is not so much different from 1.

We shall present the calculation of all contributions of diagrams
of Fig.~4(b) in our forthcoming paper.

\subsection{Calculation of the main diagram contribution
$\lambda_S$.\\
Formfactor-like cut off}

Taking into account that we have 12 different states of the top
(or anti-top) quark in the bound state $S$ (top-quark or
anti-top-quark with 3 colors and 2 spin-1/2 states), we obtain the
following integral corresponding to the diagram of Fig.~4(a):

\be \lambda_S \simeq G_{HSS}^4\int \frac{d^4q}{(2\pi )^4}
\frac{\mathfrak{F}_0\left(q^2\right)
\mathfrak{F}_0\left(\left(q+p_1\right)^2\right)
\mathfrak{F}_0\left(\left(q+p_1-p_3\right)^2\right)
 \mathfrak{F}_0\left(\left(q-p_2\right)^2\right)}{\left(q^2 - m_S^2\right)
\left(\left(q + p_1\right)^2 - m_S^2\right)\left(\left(q + p_1 - p_3\right)^2 -
m_S^2\right)\left(\left(q - p_2\right)^2 - m_S^2\right)}, \lb{21a} \ee
where $q$ is the loop 4-momentum, and $p_i$  $(i=1,2,3,4)$ are
the external 4-momenta of the Higgs bosons. Here $p_1 + p_2 = p_3
+ p_4$. The coupling constant between the three scalar particles
SSH (see \ct{DIP}) is:
\be G_{HSS} = 24 m_S \frac{g_t}{\sqrt 2}. \lb{22a} \ee
Eq.~(\ref{21a}) contains formfactors
$\mathfrak{F}_0\big(q_i^2\big)$, which describe formfactor-like
cut off.

If NBS has a very small radius then in loop calculations we can
treat such a bound state as a ``fundamental'' particle. This means
that we introduce more Feynman rules with the bound state
corresponding propagators or some appropriate effective coupling
vertices. But if we have a bound state with an extension, or the
radius, $r_0$, which is not sufficiently small, then we expect to
have in loops formfactors behaving like
\be {\mathfrak F}_0\left(q^2\right) = \exp \left(\frac 16 \left\langle \vec r^2\right\rangle q^2\right),
\lb{23a} \ee
where $q$ is the four momentum relevant for the effective vertex
or propagator in question. After the Wick rotation the loop
four-momentum $q$ will be space-like so that $q^2 < 0$. Thus, when
we want to estimate, what correction we need to include for a
given loop with an aim to improve a contribution of bound states
running around the loop, then we suppose that the exponential
formfactor (\ref{23a}), once for each propagator, comes in
multiplying the integrand of the Feynman diagram which damps the
contribution of the diagram numerically.

In Eq.~(\ref{23a}) we have for our $\lambda_S$-loop calculation:
\be \left\langle \vec r^2\right\rangle =3r_0^2. \lb{24a} \ee
Let us present now the radius of the bound state $S$ as
\be r_0 = \frac{b}{m_t}. \lb{25a} \ee
If $r_0$ is the radius $a_B$ of the Bohr Hydrogen-atom-like bound
state $S$ containing 12 top-quarks and having $Z=11/2$, then $r_0
= a_B$.

The next Sections are devoted to the correct estimation of the
radius of the new bound state $S$ of $6t + 6\bar t$ in details,
and then we compare it with the radius needed for exact MPP.

Let us continue now the calculation of the contribution
$\lambda_S$ given by diagram of Fig.~4(a).

Neglecting 4-momenta $p_i\, (i=1,2,3,4)$ in the integrand of
Eq.~(\ref{21a}) as very small, we obtain:
\be \lambda_S \approx G_{HSS}^4\int \frac{d^4q}{(2\pi
)^4}\frac{\left({\mathfrak F}_0\left(q^2\right)\right)^4}{\left(q^2 - m_S^2\right)^4}.
\lb{27a} \ee
Considering the Wick rotation, we obtain the loop four-momentum
$q_E$ which is space-like and $q^2 = - q_E^2$. Then the formfactor
(\ref{23a}) is:
\be {\mathfrak F}_0(q_E^2) = \exp \left( - \frac 12 r_0^2 q_E^2\right),
 \lb{28a} \ee
Using the expression (\ref{28a}) for Eq.~(\ref{27a}), we obtain:
$$ \lambda_S \approx \frac{G_{HSS}^4}{(2\pi )^4}\int d^4q \frac{\left(\exp \left(
\frac 12r_0^2q^2\right)\right)^4}{\left(q^2 - m_S^2\right)^4} = \frac{G_{HSS}^4}{(2\pi
)^4}\int d^4q_E \frac{\exp \left( - 2r_0^2q_E^2\right)}{\left(q_E^2 + m_S^2\right)^4}$$ $$=
\frac{G_{HSS}^4}{16\pi^2} \int\limits_0^{\infty} q_E^2dq_E^2 \frac{\exp \left(
- 2r_0^2q_E^2\right)}{\left(q_E^2 + m_S^2\right)^4} = \frac{G_{HSS}^4}{16\pi^2
m_S^8}\int\limits_0^{\infty} q_E^2dq_E^2\frac{\exp \left( -
2r_0^2q_E^2\right)}{\left(q_E^2/m_S^2 + 1\right)^4}$$ $$=
\left(\frac{G_{HSS}}{2m_S}\right)^4 \frac{1}{\pi^2 }
\int\limits_0^{\infty} ydy \frac{\exp\left(-\left(2r_0^2m_S^2\right)y\right)}{(y+1)^4} $$
\be\approx \frac{1}{4\pi^2}\left(\frac{G_{HSS}}{r_0m_S^2}\right)^4
= \frac{1}{\pi^2}\left( \frac{6g_t}{b}\cdot
\frac{m_t}{m_S}\right)^4,
 \lb{30a} \ee
where $y=q_E^2/m_S^2$ and $G_{HSS}$ is given by Eq.~(\ref{22a}).

If any $S$-resonance (with $m_S\approx 300$ GeV or $750$ GeV)
gives
\be
 \lambda_S \simeq 0.01, \lb{31a} \ee
then this contribution transforms the metastable (blue) curve of
Fig.~3 into stable (red) curve, and we have exact vacuum stability
and exact MPP.

From Eq.~(\ref{30a}) we have:
\be \lambda_S \simeq \frac{1}{\pi^2}\left(
\frac{6g_t\xi}{b}\right)^4, \lb{32a} \ee
where $$\xi=\frac{m_t}{m_S}.$$

Vacuum stability corresponds to the condition:
\be \left(\frac{\xi}{b}\right)^4 \simeq
0.01\frac{\pi^2}{\left(6g_t\right)^4}\approx 0.00008. \lb{33ab} \ee
or
\be \frac{\xi}{b} \simeq 0.095, \lb{34ab} \ee
what gives:
\be b\simeq \frac{\xi}{0.095}. \lb{35ab} \ee
We see that $\xi_1\approx 173/750 \approx 0.231$ for the resonance
with the mass $m_S\approx 750$ GeV and $\xi_2\approx 173/300
\approx 0.577$ for $m_S\approx 300$ GeV. Very likely one of the
two mentioned resonances could turn out to be a statistical
fluctuation. Therefore, if only the resonance with mass 750 GeV
provides the vacuum stability, then its radius $r_0=b/m_t$ has to
have:
\be b = b_1\simeq \frac{0.231}{0.095}\approx 2.34. \lb{36ab}
\ee
But in the case of the 300 GeV resonance (only), the vacuum
stability is possible for
\be b = b_2\simeq \frac{0.577}{0.095}\approx 6.07. \lb{37ab}
\ee
If both resonances with different $b'_i$ $(i=1,2)$ give their
contribution to $\lambda_S$, then we must in order to get the
exact MPP have:
\be \left(\frac{\xi_1}{b'_1}\right)^4 +
\left(\frac{\xi_2}{b'_2}\right)^4 \simeq 0.01\frac{\pi^2}{\left(6g_t\right)^4}
\approx 0.00008. \lb{38ab} \ee
In Section 6 we obtained $b\simeq 2.34$ (see below), for the
experimental values $g_t\approx 0.935$ and $m_S\simeq 750$ GeV,
it gives:
\be
 \lambda_S \simeq 0.009, \lb{31a} \ee
what transforms the metastable (blue) curve of Fig.~3 into stable
(red) curve. Of course, the uncertainty coming from the
contributions of diagrams shown in Fig.~4(b) can reach 25\%, and
then we have:
\be
 \lambda_S \simeq 0.009 \pm 0.002. \lb{32a} \ee
It is not excluded that the mentioned resonance with the mass
$m_S\approx 300$ GeV is not a bound state $6t + 6\bar t$, but only
a statistical fluctuation.

It is very important to investigate experimentally the radii of
the discovered resonances.

\section{Estimation of the radius of the new bound state S of $6t +
6\bar t$ in the mean field approximation}

The kinetic energy term of the Higgs field and the top-quark
Yukawa interaction are given by the following Lagrangian density:
\be L = \frac 12 {\cal D}_{\mu}\Phi_H {\cal D}^{\mu}\Phi_H
 + \frac{g_t}{\sqrt 2}\bar \psi_{tR}\psi_{tL}\Phi_H. \lb{33a} \ee
According to the Salam-Weinberg theory, the top-quark mass $M_t$
and the Higgs mass $M_H$ are given by relations (\ref{11a}) and
(\ref{12a}) with $v\approx 246$ GeV.

Let us imagine now that the NBS is a bubble in the EW-vacuum and
contains $N$ top-like constituents. It is known that inside the
bubble (bag) the Higgs field can modify its VEV. Implications
related with this phenomenon have been discussed in
Refs.~\ct{2nbs,3nbs} and Refs.~\ct{Bag1,Bag2,Bag3}. Then inside
bound state $S$ the VEV of the Higgs field is smaller than v:
\be v_0 = \left\langle 0|\Phi_H|0 \right\rangle , \quad {\rm{where}}
 \quad \frac{v_0}{v} < 1, \lb{34a} \ee
and the effective Higgs mass inside the bubble/bag are smaller
than the corresponding experimental masses: $m_h = (v_0/v) M_H$ .
In this case the attraction between the two top (or anti-top)
quarks is presented by the Yukawa type of potential:
\be V (r) = - \frac{g^2_t /2}{4\pi r}\exp\left( - m_h r\right). \lb{35a}
\ee
Assuming that the radius $r_0$ of the bubble is small:
\be m_hr_0 \ll 1, \lb{36a} \ee
we obtain the Coulomb-like potential:
\be V (r) = - \frac{g^2_t /2}{4\pi r}. \lb{37a} \ee
The attraction between any pairs $tt, \,t\bar t, \,\bar t\bar t$
is described by the same potential (\ref{37a}). By analogy with
Bohr Hydrogen-atom-like model, the binding energy of a single
top-quark relatively to the nucleus containing $Z = N/2$
top-quarks have been estimated in Refs.~\ct{1nbs,2nbs,3nbs,4nbs}.
The total potential energy for the NBS with $N = 12$ has $Z=11/2$
(see explanation in Ref.~\ct{4nbs}) and is:
\be V (r) = - \frac{11}{2}\frac{g^2_t /2}{4\pi r}. \lb{38a} \ee
Here we would like to comment that the value of the mass $m_h$,
which belongs to the Higgs field inside the NBS $6t+ 6\bar t$, can
just coincide with estimates given by
Refs.~\ct{Kuch1,Kuch2,Kuch3,Rich}. The results: ${\rm{\bf
max}}(m_h) = 29$ GeV and ${\rm{\bf max}}(m_h) = 31$ GeV correspond
to Ref.~\ct{Kuch2} and Ref.~\ct{Rich}, respectively.

In Refs.~\ct{1nbs,2nbs,3nbs,4nbs} it was suggested that the Higgs
boson, responsible for generating the masses of fermions in the
Standard model, couples more strongly to the heavy quarks (top,
beauty, etc.,) than to the light ones. The question has been
raised whether the attraction mediated by Higgs exchange could
produce new type of bound states. If the Higgs boson coupling to
top quarks is large enough to generate a whole spectroscopy of new
bound states (NBS), then for large enough number of quarks this
binding energy can be strong enough to stabilize top quarks, as
bound neutrons are stabilized in nuclei. The particular state
discussed in \ct{1nbs,2nbs,3nbs,4nbs} is for the number of quarks
N = 12, 6 t-quarks and 6 anti-t-quarks, with all spin and color
values, which all occupy the same lowest 1S orbital. In a simple
approximation the binding energy for this state is so large that
the total mass of this 1S bound state turns to be zero (with
claimed accuracy). At least, the mass of any NBS is much smaller
than the mass of 12 top-quarks. Refs.~\ct{1nbs,2nbs,3nbs,4nbs}
considered proposals of how to find such states at Tevatron and
LHC.

In the Standard model the interaction of t-quarks with the Higgs
boson is proportional to the large mass of the t-quark, $M_t =
173.34$ GeV \ct{PDG}, with a coupling $g_t = 0.935 \pm 0.008$
\ct{PDG}. Therefore the effective Coulomb coupling is about as
strong as a QCD coupling constant $\alpha_s$ at the same scale,
with the additional advantage that quarks and anti-quarks of any
color are equally attracted by the Higgs exchange. Thus the
binding energy should grow with the number of quarks.

The calculations of Refs.~\ct{1nbs,2nbs,3nbs,4nbs} gave only a
rough estimate for the binding energy, which was based on analogy
with Bohr energies of the Hydrogen atom. However, the Bohr atom
has an attractive Coulomb center, while for multi-top bound states
theory considers the charge which is smeared out over the volume
of the multi-top bound state. The calculations of
Refs.~\ct{Kuch1,Kuch2,Kuch3,Rich} show that this leads to the huge
difference for 12 t-quarks, which are not deeply bound at all, but
rather unbound, if realistic limits on the Higgs mass is used.

The authors of Refs.~\ct{Kuch1,Kuch2,Kuch3,Rich} have used first
the variational approach, and then a self-consistent mean-field
approach. The many-body and recoil corrections are expected to be
small, $\simeq 1/N$. The number of t-quarks $N = 12$ gives the
accuracy of the binding energy $\simeq 10\%$. For weak coupling
the non-relativistic approximation is valid. Then the interaction
between top quarks due to Higgs boson exchange may be described by
the following Hamiltonian (in units $\hbar=c=1$):
\be
 H = \sum_{i=1}^N \frac{\vec p^2}{2m_t} + \sum_{i <
 j}V\left(\vec r_{ij}\right), \lb{39a} \ee
where $r_{ij}$ are the interquark distances, and
\be V (r) = - \frac{\alpha_H}{r}\exp\left( - m_h r\right). \lb{40a} \ee
Here $m_h$ is the Higgs mass inside the bound state, and the
coefficient $\alpha_H$ is introduced for the strength of the Higgs
boson exchange. The t-channel exchange in the top-quark scattering
leads to the effective Coulomb coupling $\alpha_H = g_t^2/(8\pi)$.
The inclusion of s- and u- channels makes the effective
interaction between t-quarks more strong (see \ct{4nbs}). In
general:
\be \alpha_H = \frac {{\left(g_{t}\right)}_{eff}^2}{8\pi} = \frac{\kappa
g_t^2}{8\pi}. \lb{41a} \ee
The authors of Refs.~\ct{Kuch1,Rich} presumed that $\kappa\approx
2$ and $\alpha_H \approx g_t^2/(4\pi)$. Considering $g_t\approx
1$, they obtained $\alpha_H \approx 1/(4\pi)\approx 0.08$, and
used this value of $\alpha_H$ in their calculations.
For a preliminary investigation of the existence of new type of
bound states due to the Higgs exchanges, the authors of
Refs.~\ct{Kuch1,Rich} have used the Hamiltonian (\ref{39a}) and
examine its spectral properties. If N, the number of constituents,
does not exceed 6 top quarks and 6 top anti-quarks, the colour and
spin degrees of freedom can endorse the constraints of
anti-symmetrisation, and for the orbital variables, the
Hamiltonian (\ref{39a}) can be considered as acting on effective
bosons. This is a reason why our attention has been focused on the
$6t+6\bar t$ system, which in Ref.~\ct{Rich} was named
``dodecatoplet'', by analogy with a ``pentaquark''.

The level energies $E_N$ of the Hamiltonian $H$, and in particular
its ground-state, are given by the following scaling way:
\be E_N = \left(m_t, \alpha_H, m_h\right) =
\frac{m_h}{m_t}\epsilon_N(G), \lb{42a} \ee
where
\be G = \frac{m_t}{m_h}\alpha_H. \lb{43a} \ee
The variational approach assumes that the wave function of the
multi-top system incorporates a product of the N orbitals. The
ground-state energy $\epsilon_2 = E_2$ has the variational upper
bound:
\be \tilde{\epsilon}_2 = {\rm{min}}_a\left[t(a) - Gp(a)\right], \lb{44a}
\ee
corresponding to a single normalized Gaussian:
\be \psi_a(r)\propto \exp\left(-\frac{a}{2}r^2\right), \lb{45a} \ee
where the range parameter $a$ (dimensionless in units $m_h=1$) is
optimized.

The radial wave function $\phi_a(r) = r\psi_a(r)$ gives the
dimensionless radius of the bound state:
\be {\tilde r}_0 = \frac{1}{2\sqrt a} = b\frac{m_h}{m_t}.
 \lb{45b} \ee
In Eq.~(\ref{44a}) we have functions of the radius of toponium -
the bound state with N=2:
\be t(a) = \frac{3a}2, \quad {\rm{and}} \quad p(a) = 2\sqrt
{\frac{a}{\pi}} - \exp\left(\frac 1{4a}\right)erfc\left(\frac 1{2\sqrt
a}\right).
 \lb{46a} \ee
For $N$ constituents the variational energy is:
\be \tilde{\epsilon}_N = {\rm{min}}_a(N - 1)\left[t(a) - N G p(a)\right].
 \lb{47a} \ee
Constructing a function:
\be \frac{\tilde{\epsilon}_N}{N-1} = {\rm{min}}_a\left[t(a) - N G p(a)\right]
= {\rm{min}}_aF(z), \lb{48a} \ee
where $F(z)$ depends on the variable $z=1/(2\sqrt a)={\tilde r}_0
$, and using functions (\ref{46a}), we obtain for $N=12$ the
following expression:
\be F(z) = t(a) - 6 G p(a) = \frac{3}{8 z^2} - \frac{6 G}{\sqrt \pi
z} + 6 G e^{z^2} erfc\left(z\right). \lb{49a} \ee
Minimization of Eq.~(\ref{49a}) gives the following equation:
\be \frac{1}{2 G\left(z_0\right)} = \frac{4}{\sqrt \pi} z_0 - \frac{8}{\sqrt
\pi} z_0^3 + 8 z_0^4 e^{{z_0}^2} erfc\left(z_0\right) =
f\left(z_0\right), \lb{50a} \ee
where
\be f(z) = \frac{4}{\sqrt \pi} z - \frac{8}{\sqrt \pi} z^3 +
8z^4 e^{{z}^2} erfc\left(z\right). \lb{50b} \ee
Here $z_0$ is a position of the minimum of the binding energy,
where $F'(z)|_{z=z_0}=0$. The function (\ref{50b}) has a maximum
at $1/2G\approx 1.19$ and $z_0\approx 1.3$. According to
Eq.~(\ref{43a}), we have a maximal value of $m_h$:
$m_h({\rm{max}}) = 31$ GeV \ct{Rich} for $\alpha_H\approx 0.075$,
what is close to $\kappa =2$ and $g_t = 1$ in Eq.~(\ref{41a})
\ct{Rich}. The result max $m_h = 29$ GeV was obtained in
Ref.~\ct{Kuch1}. We see that 12 top-quarks cannot bind, if the
Higgs effective mass inside the bound state is larger than
$m_h({\rm{max}})$: binding energy per quark for $S$,
$\epsilon_{12}/11$, is negative for $G > G_{min} =
1/2f(z_0)\approx 0.42$ (compare with the result of
Ref.~\ct{Rich}).

In Eqs.~(\ref{49a}) and (\ref{50a}) we have:
\be G = \gamma \alpha_H \quad {\rm{and}}\quad z =
\frac{b}{\gamma}, \lb{51a} \ee
where $\gamma = m_t/m_h$, and
\be \gamma_{min}\approx \frac {2G_{min}}{2\alpha_H}=
 \frac{1}{2.38\alpha_H}\approx 5.6 \lb{52a} \ee
for $\alpha_H\approx 0.075$. As a result, we have:
\be
 b\left({\rm{max}}\; {m_h}\right)\approx 1.3\gamma_{min}\approx 1.3\cdot 5.6 = 7.28. \lb{53a} \ee
This is a result for $b$, if the effective Higgs mass inside the
bound state is maximal: $m_h({\rm{max}}) \approx 31$ GeV
\ct{Rich}. However, we do not believe that $m_h$ inside the bound
state is large.

For $m_h = 0$ ($z = 0$) we have obtained (from (\ref{50b}) and
(\ref{51a})) the following result:
\be
 \left.\frac {1}{b\alpha_H} = \frac {f(z)}{z}\right|_{z=0} = \frac{8}{\sqrt \pi}, \lb{54a} \ee
and
\be b\approx 2.95. \lb{55a} \ee
Below, in contrast to the mean field approximation, developed in
Refs.~\ct{Kuch1,Kuch2,Kuch3,Rich}, we give an alternative
estimation of radius of the bound state $S$.

\section{An alternative radius estimation for the new bound state
$S$}

Here we have an aim to compute the radius of the bound state $S$
by an alternative way presumably described in Ref.~\ct{4nbs}.

\subsection{The ``eaten Higgs'' exchange corrections}

Now we want, contrary to what was considered by Froggatt and
Nielsen in Ref.~\ct{4nbs}, to take into account the ``eaten
Higgs''- corrections which we called ``local''. Really we follow
the thoughts of \ct{4nbs} that the inclusion of the ``eaten Higgs''
exchange leads to an effect as if the $g_t$ had in first
approximation a factor $4^{1/4}$ (then $\kappa=2$ in
Eq.~(\ref{41a})). But by more careful estimation, we rather argued
that instead of just this factor $4^{1/4}=\sqrt 2$, we should use
$\sqrt 2\exp(- 4.0 \%) = 1.359$. Now because of the existence of
the gluon contribution included into the definition of
$\alpha_H$, this $\alpha_H$ is not quite increased by a factor
$\kappa \approx 1.359^2\approx 1.85$ due to such an effect. Rather
we might take into account that for the binding energy in the Bohr
atom approximation, instead of $(1/2) \alpha_H^2$, which is
$\propto g_t^4$ in the case of ignoring of gluon interactions, we
have an effective power $g_t^{(2.9)}$. This 2.9 is estimated by
saying that the gluon term in the $\alpha_H$ makes up about
$27.4\% $ of the total $1.83 + 4g_t^2= 1.83 + 4.84$ (see
\ct{4nbs}). Thus, the Higgs part of $\alpha_H$ is 72.6 \%, and an
effective power dependence is $(1.359 g_t)^{(2.9)} = 2.434
g_t^{(2.9)}$, what means that the binding energy is changed by a
factor $1.359^{(2.90)} = 2.434$.

Now the philosophy of the ``eaten Higgs'' correction is implemented
by a change in the binding energy without it coming from ``long
distance'' calculation by Richard \ct{Rich}. The effect of the
simply changing the binding energy is in writing the binding
energy per quark as $\epsilon_{(12)}/11 = Am_t/m_h$ and
considering the $A$ being increased by the factor 2.434. Thus, if
for the purpose of achieving the ``light'' mass (much lighter than
$12 M_t\approx 2$ TeV) for the bound state $S$, we need $A=0.49$,
then we should require that {\em before making the ``eaten Higgs''
correction} the value of $A$ should be decreased by a factor
2.434, that is, we should require that an estimate of the binding
energy ignoring the ``eaten Higgs'' interaction should only lead to
the binding energy $\epsilon_{(12)}/11 = A_{lbe}m_t/m_h$,
where $A_{lbe}$ is:
\be A_{lbe} = A_{\hbox{(before adding local binding energy)}} =
\frac{0.49}{2.434} = 0.201. \lb{1w} \ee

\subsection{Froggatt-Nielsen corrections to the $S$-radius estimates}

In Ref.~\ct{4nbs} Froggatt and Nielsen estimated the radius of the
bound state $S$, using the assumption that the binding energy per
quark is given by the value $Am_t = m_t/2$. Here $A = 0.5$ is
needed for obtaining the light mass for the bound state $S$. But
now formally instead of $A = 0.5$ we shall use $A_{lbe} = 0.201$
given by Eq.~(\ref{1w}). With this new value of the assumed
binding energy we shall follow the virial theorem which was used
previously in Ref.~\ct{4nbs} for $S$-radius estimates.

This means that if we, for example, imagine that a given quark (or
anti-quark) is so close to another one that the potential between
them dominate and is proportional to $1/r$ (here we ignore the
Higgs mass, considering its correction later), then from the
virial theorem we would have that in the ground state of this
two-quark system the average potential energy $\left\langle V\right\rangle $ is given by
the following equality:
\be \left\langle V\right\rangle = -2\left\langle T\right\rangle , \lb{2w} \ee
where $\left\langle T\right\rangle $ is the average kinetic energy. Then the binding energy
$\epsilon_{(12)}/11$, apart from the sign, coincides with just the
kinetic energy:
\be \epsilon_{(12)}/11 = \left\langle V\right\rangle - \left\langle T\right\rangle = - \left\langle T\right\rangle . \lb{3w} \ee
In general, we have:
\be \left\langle \vec{p}^2\right\rangle \approx 2m_{\hbox{t (eff.rel.)}}\left\langle T\right\rangle , \lb{5w} \ee
where $m_{\hbox{t (eff.rel.)}}$ is the effective relativistic mass
of the top-quark inside the bound state. Of course, this relation
is true at least for transverse (extra) momenta. Here the ``eff''
means that the rest top-quark mass is not to be given as the usual
$m_t$ but it is obtained by an average Higgs field value in the
interior of the bound state.

In the previous article \ct{4nbs} it was assumed (see again the
discussion of this problem below) that as an approximation, the
Heisenberg uncertainty relation $\sigma_x\sigma_{p_x}\ge 1/2$ has
{\em equality} rather than {\em inequality}. That is, we assume
that the wave functions can be approximated by the following
relations:
\begin{align}
\left\langle x^2\right\rangle \left\langle p_x^2\right\rangle &=1/4,\\
\left\langle \vec{r}^2\right\rangle \left\langle \vec{p}^2\right\rangle &=9/4.
\end{align}
If these relations indeed were true, then we would get:
\be \left\langle \vec{r}^2\right\rangle =\frac{9}{4\left\langle \vec{p}^2\right\rangle } = \frac{9}{8 m_{\hbox{t
(eff.rel.)}}\left\langle T\right\rangle } = \frac{9}{8m_{\hbox{t (eff.rel.)}}|A|m_t}
=\frac{9}{8(1-|A|)|A| m_t^2}. \lb{7w} \ee
This means (in the notations used) that
\begin{eqnarray}
\left\langle \vec{r}^2\right\rangle & =& 3r_0^2,\\
r_0 &=& \frac{b}{m_t},
\end{eqnarray}
and we obtain:
\be b=\sqrt{\frac{\left\langle \vec{r}^2\right\rangle }{3}} m_t = \frac{3}{8(1-|A|)|A|} =
\frac{3}{8(1-0.201)0.201} = 2.34, \lb{9w} \ee
when we have inserted for $|A|$ the value 0.201, taking into
account the eaten Higgs correction (\ref{1w}).

\section{Summary and Conclusion}

In this paper we assumed that recently discovered at the LHC new
resonances with masses $m_S\approx 300$ GeV and $750$ GeV, or just
one of them, are new scalar $S$ bounds states $6t + 6\bar t$,
earlier predicted by C.D.~Froggatt and authors. It was shown that
these NBS can provide the vacuum stability and exact accuracy of
the Multiple Point Principle, according to which the two vacua
(existing at the Electroweak and Planck scales) are degenerate.

We discussed the possibility of different new bound states (LHC
resonances) to give the correction to the Higgs mass coming from
the bound states of 6 top and 6 anti-top quarks. We showed that
the value of their radii are essential for the transformation of
the metastable SM vacuum into the stable one.

We calculated the main contribution of the $S$-resonance to the
renormalization group evolution of the Higgs quartic coupling
$\lambda$, and showed that the resonance with mass $m_S\approx
750$ GeV, having the radius $r_0 = b/m_t$ with $b\approx 2.34$,
gives the positive contribution to $\lambda$, equal to the
$\lambda_S\approx + 0.01$. This contribution compensates the
asymptotic value of the $\lambda\approx - 0.01$, which was earlier
obtained in Ref.~\ct{Deg}, and therefore transforms the
metastability of the EW vacuum into the stability.

We have considered the calculation of the NBS radius in the model
of the mean field approximation, which was developed by authors of
Refs.~\ct{Kuch1,Kuch2,Kuch3,Rich}. But then we considered an
alternative way of the radius calculation for the $S$ bound state,
developed in the Froggatt-Nielsen relativistic model. The last
model gave successful results for our aims to get the SM vacuum
stability and an almost exact effect of the MPP.

\section{Acknowledgments}
We deeply thank Colin D. Froggatt for the first noticing of the
750 GeV peak and its supposed identification with the bound state.
In fact, we started working with the 300 GeV one as the presumed
candidate.

H.B.N. wishes to thank the Niels Bohr Institute for the status of
professor emeritus and corresponding support. L.V.L. greatly
thanks the Niels Bohr Institute for hospitality and Prof. Holger
Bech Nielsen for the financial support and very fruitful
collaboration. C.R.D acknowledges the JINR and greatly thanks
Prof. D.I. Kazakov for support.

\begin{figure}[H]
\centering
\includegraphics[scale=0.1235]{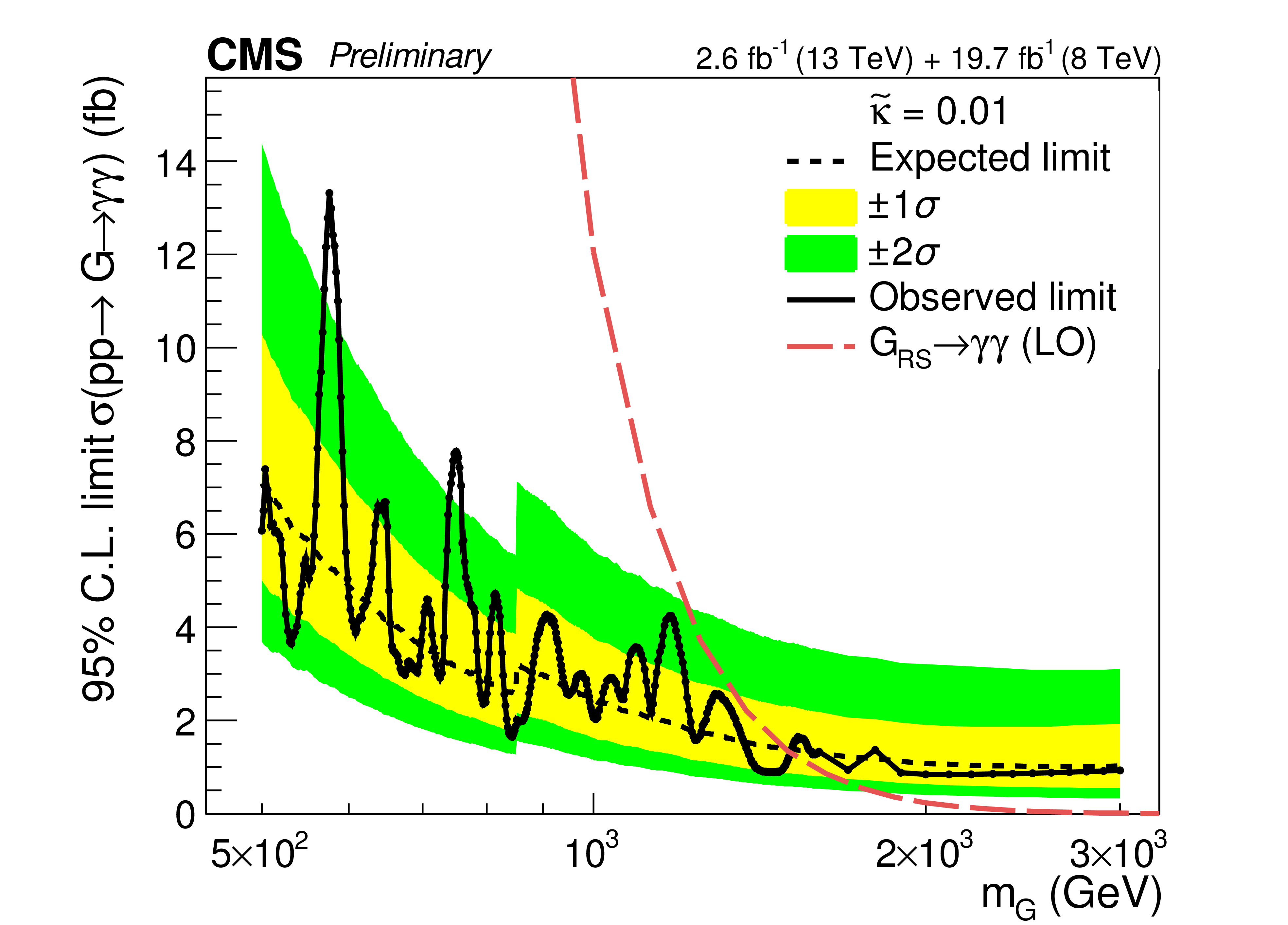}
\par
\centering \vspace{0.5cm}
\includegraphics[scale=0.23]{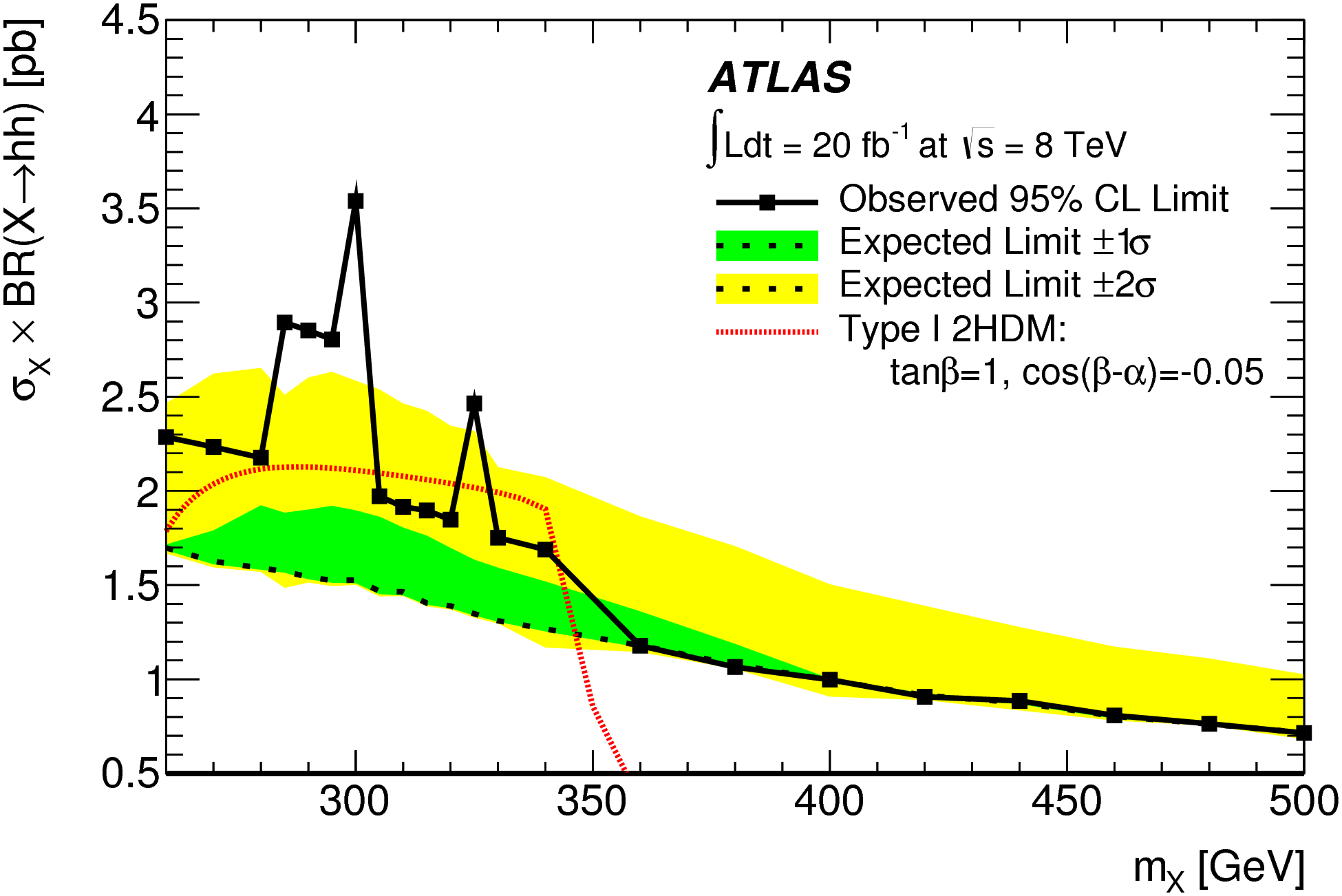}
\caption
{The first figure presents searches for a new physics in high mass
diphoton events in proton-proton collisions at 13 TeV and 8 TeV combined analysis. ATLAS and
CMS Collaborations show a new resonance in the diphoton
distribution at an invariant mass of 750-760 GeV. The next figure
presents searches for resonant and non-resonant Higgs boson pair
production using 20.3 ${\rm{ fb}^{-1}}$ of proton-proton
collision data at $\sqrt s = 8$ TeV generated by the LHC and
recorded by the ATLAS detector in 2012. The results show a
resonance with mass $\approx 300$ GeV}
\end{figure}

\begin{figure}[H]
\centering
\includegraphics[scale=0.65]{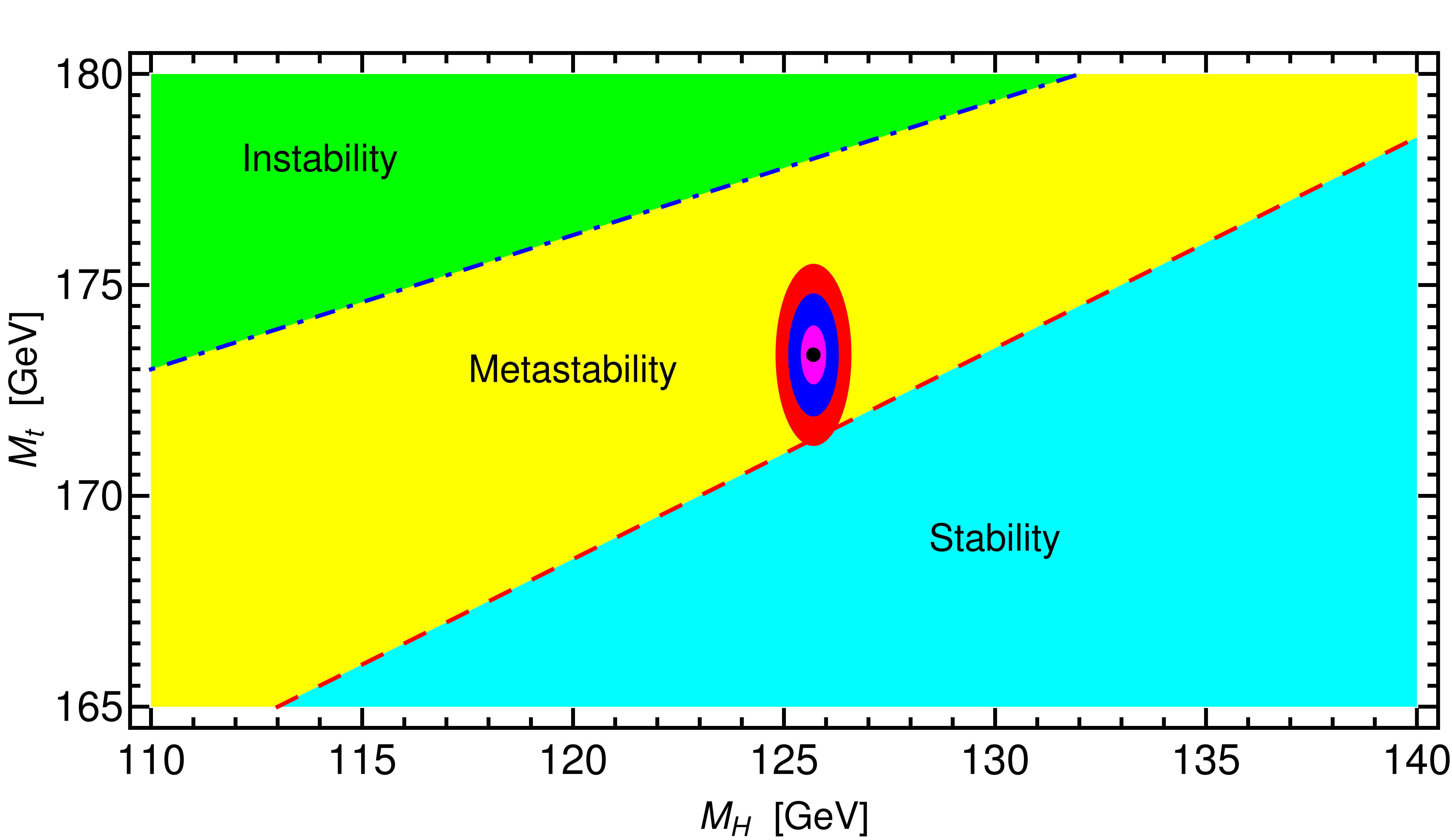}
\caption
{The stability phase diagram obtained according to the standard
analysis. The $(M_H,M_t)$ plane is divided in three sectors:
absolute stability, metastability and instability regions. The dot
indicates current experimental values $M_H\simeq 125.7$ GeV and
$M_t\simeq 173.34$ GeV. The ellipses take into account $1\sigma,\,
2\sigma$ and $3\sigma$, according to the current experimental
errors.}
\end{figure}

\begin{figure}[H]
\centering
\includegraphics[scale=0.65]{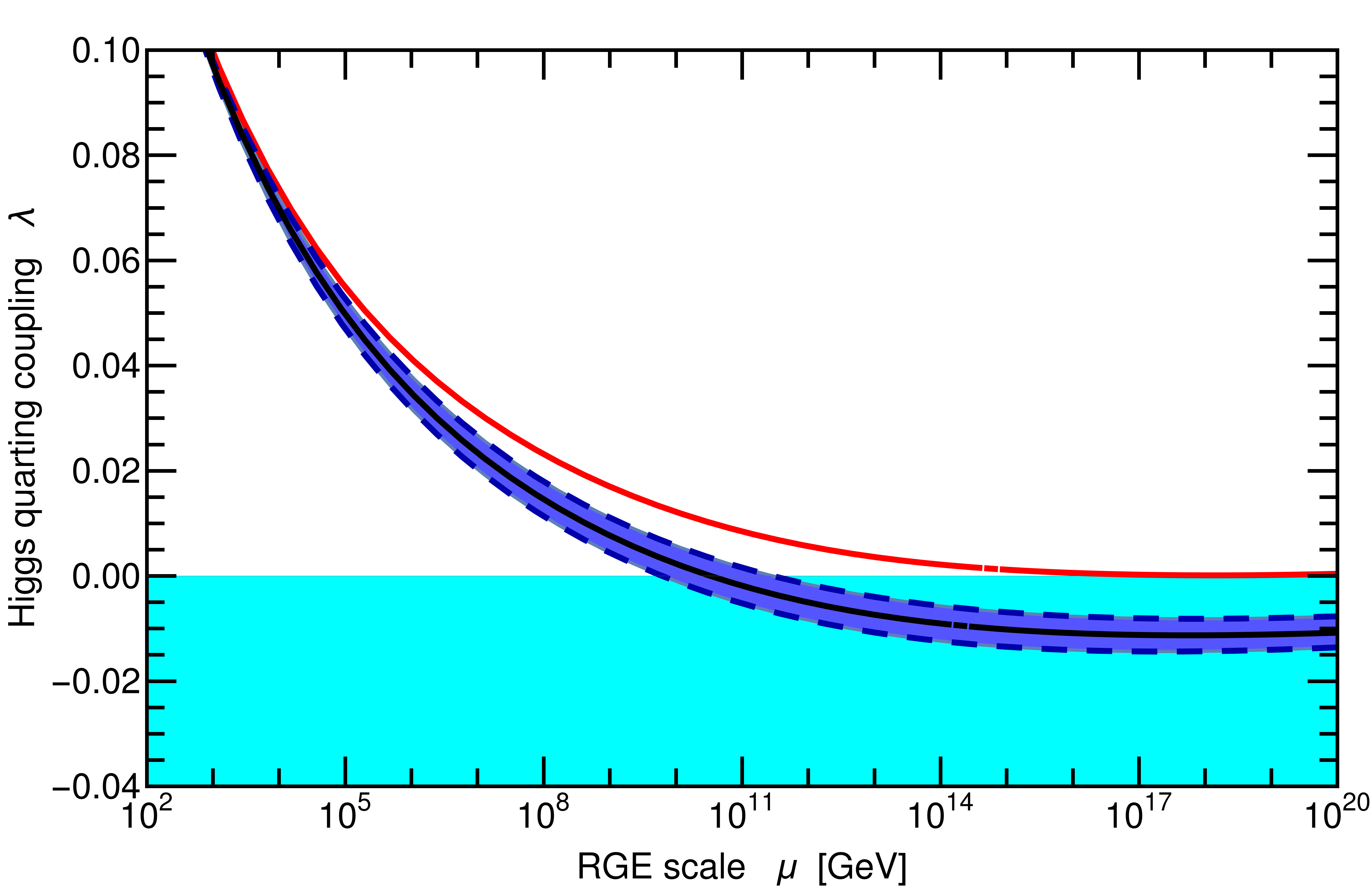}
\caption
{The RG evolution of the Higgs selfcoupling $\lambda$ for
$M_t\simeq 173.34$ GeV and $\alpha_s = 0.1184$ given by $\pm
3\sigma$. Blue lines present metastability for current
experimental data, red (thick) line corresponds to the stability
of the EW vacuum}
\end{figure}

\begin{figure}[H]
\centering
\includegraphics[scale=0.39]{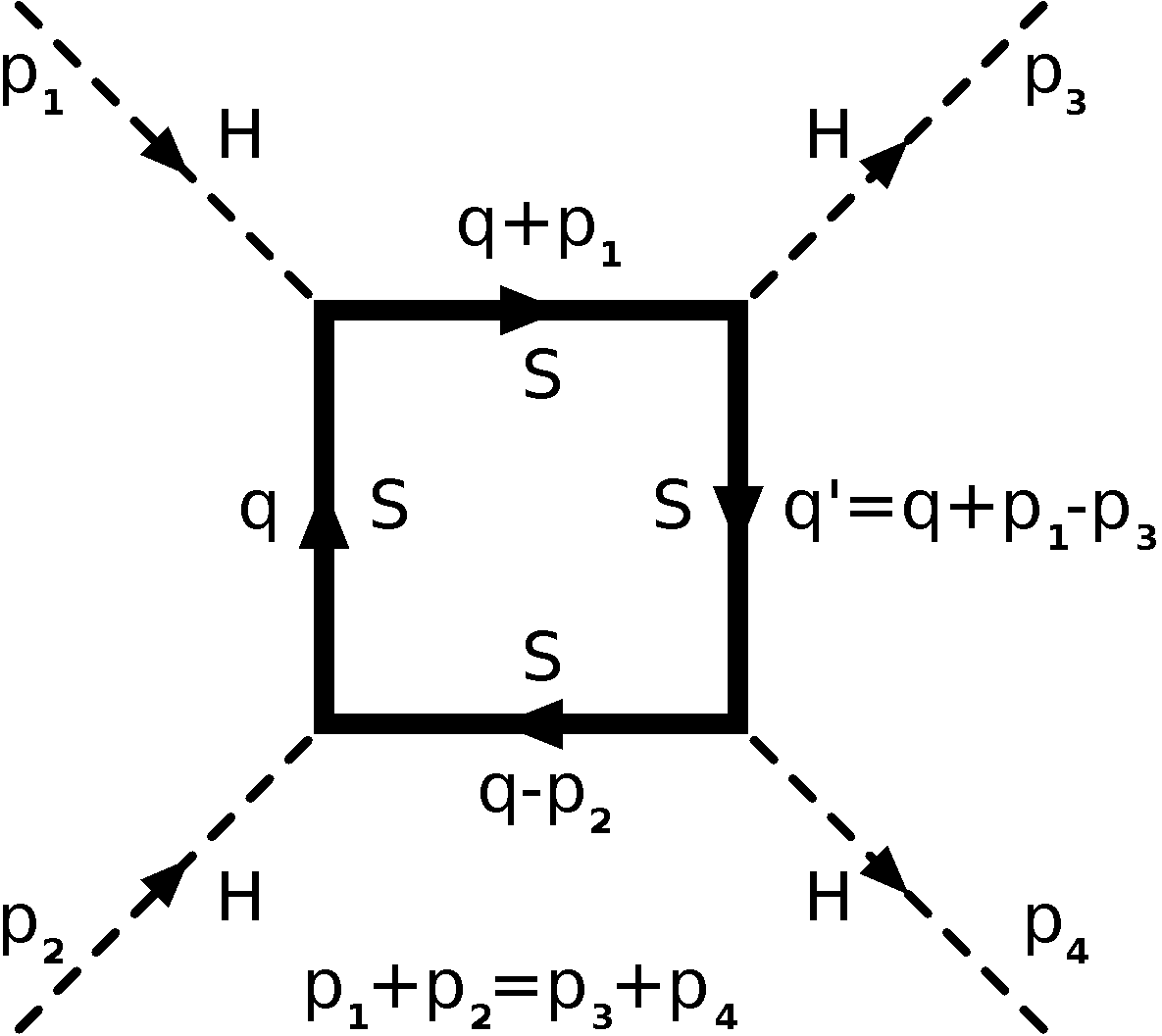}
\caption{(a) The Feynman diagram corresponding to the main
contribution of the $S$ bound state $6t + 6\bar t$ to the running
Higgs selfcoupling $\lambda$.}
\end{figure}

\setcounter{figure}{3}
\begin{figure}
\centering
\includegraphics[scale=0.28]{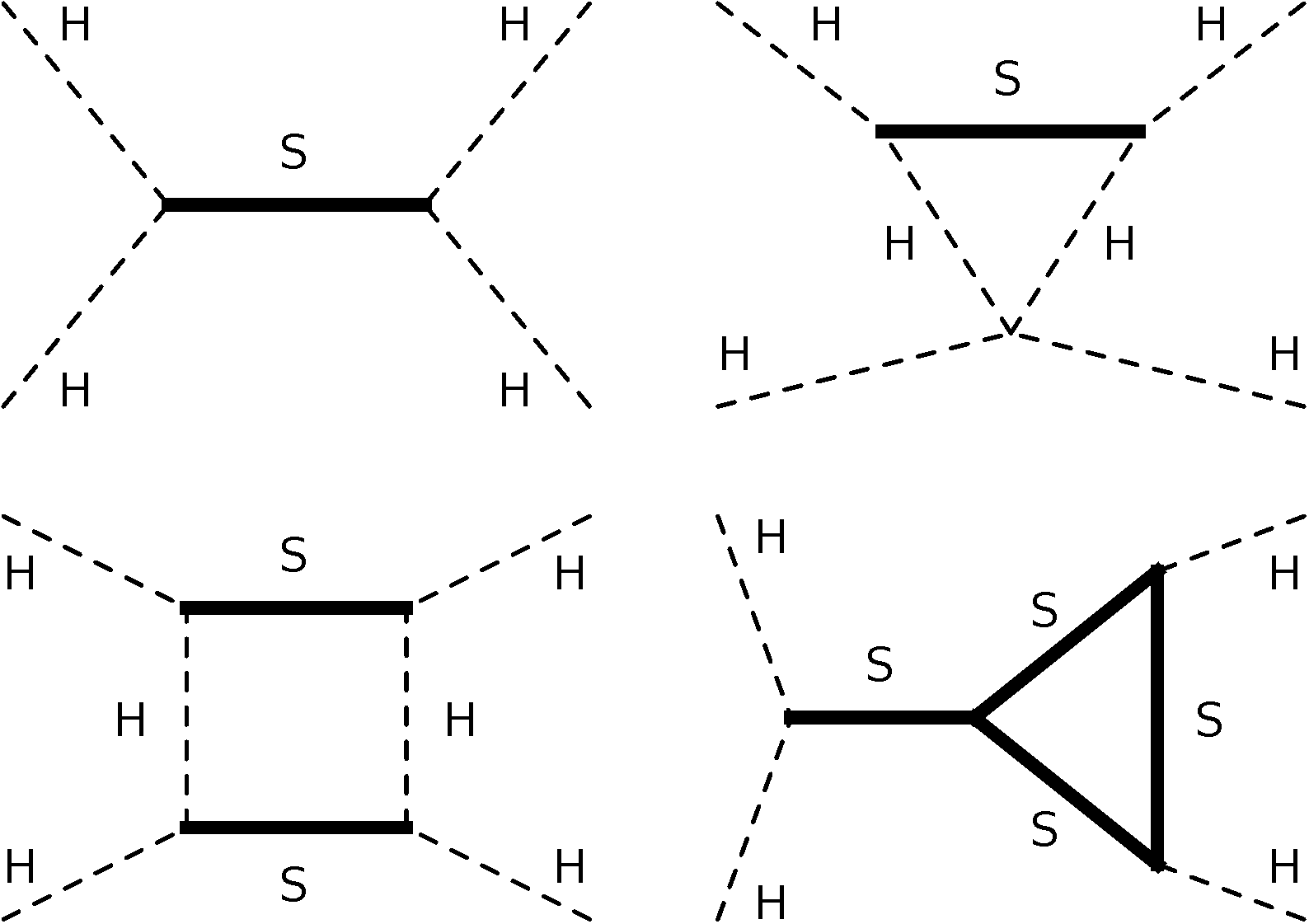}
\par
\centering \vspace{1cm}
\includegraphics[scale=0.28]{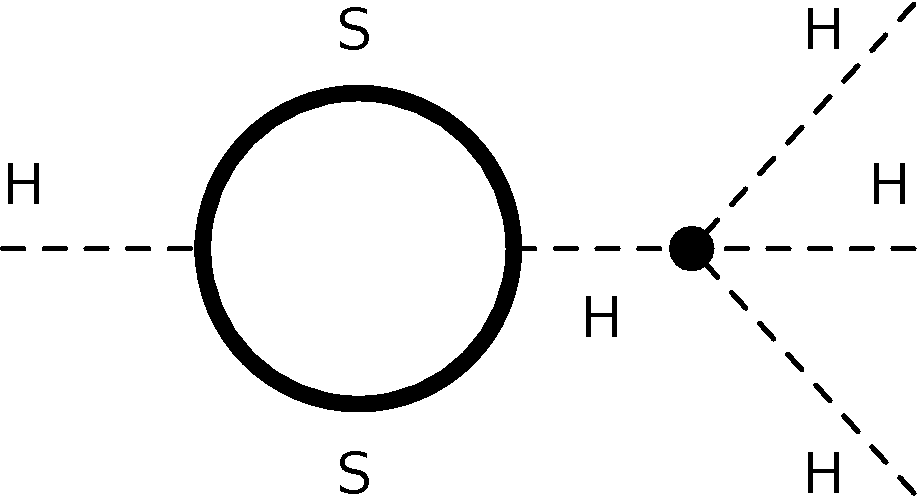}
\caption{(b) Feynman diagrams of other contributions of NBS to the
$\lambda_S$, which are smaller within 20-25\%.}
\end{figure}

\end{document}